\documentclass[prb,showpacs,preprint,floatfix]{revtex4-1}

\usepackage{amsmath}
\usepackage{amsthm}
\usepackage{algorithm}
\usepackage{algpseudocode}
\usepackage{pgfplots}
\usepackage{comment}
\usepackage{tikz}
\usepackage{subcaption}

\begin{document}

\title{Cognitive biases can move opinion dynamics from consensus to signatures of transient chaos}

\author{Emily Dong}
\affiliation{W. M. Keck Science Department, Pitzer, Scripps, and Claremont McKenna College, Claremont, California 91711}

\author{Sarah Marzen}
\email{smarzen@cmc.edu}
\affiliation{W. M. Keck Science Department, Pitzer, Scripps, and Claremont McKenna College, Claremont, California 91711}

\date{\today}
\bibliographystyle{unsrt}

\begin{abstract}
Interest in  how democracies form consensus has  increased  recently, with statistical physics and economics approaches both suggesting that there is convergence to a fixed point in belief networks, but with fluctuations in opinions when there are ``stubborn'' voters. We modify a model of opinion dynamics in which agents are fully Bayesian to account for two  cognitive biases: 
confirmation bias and in-group bias. Confirmation bias occurs when the received information is considered to be more likely when it aligns with the receiver's beliefs. In-group bias occurs when the receiver further considers the information to be more likely when the receiver's beliefs and the sender's beliefs are aligned. We find that when there are no cognitive biases, a network of agents always converges to complete consensus. With confirmation bias alone,  polarization can occur. With both biases present, consensus and polarization are  possible, but when agents attempt to counteract confirmation bias, there can be signatures of transient chaos and ongoing opinion fluctuations. Based on this simple model, we conjecture that complex opinion fluctuations might be a generic feature of opinion dynamics when agents are Bayesian with biases.
\end{abstract}


%

\maketitle

\section{Introduction}

Misinformation, disinformation, and cognitive biases can lead to dire consequences for a democracy. Anti-vaxxers, climate change deniers, and election deniers either prevent   people from taking action that would benefit the population as a whole (e.g., vaccinations and  reduction of  society's carbon footprint) or galvanize people to pursue illegal activities (e.g., storm the Capitol). It is therefore imperative that we study opinion dynamics with a goal of understanding possible interventions which could lead to the ``wisdom of the crowd'' deciding on the truth as quickly as possible, rather than mob rule based on false information.

A possible major impediment to effectively deploying the wisdom of the crowd in a democracy is cognitive bias. There are many types of cognitive biases. The two that we will consider  are confirmation bias\cite{nickerson1998confirmation} and in-group bias.\cite{hewstone2002intergroup} Agents with confirmation bias evaluate the data that already agrees with their beliefs as being more likely to be true. We can imagine that agents with confirmation bias simply confirm their initial beliefs, leading to polarization in the society.\cite{fernandes2023confirmation}
In-group bias causes an agent to more favorably treat opinions from those who are more similar to themselves. A  form of this bias can lead to individuals  to more likely associate with people with similar opinions and  lead to polarization.  These two cognitive biases harm the ability of a group of agents to decide on the truth and act accordingly.

The possibility that cognitive biases can lead to decisions inconsistent with well-confirmed  data leads to an interesting question: should agents try to counteract their cognitive biases? For instance, should they implement a form of anti-confirmation bias; that is, should    agents  evaluate the data to be more likely when it disagrees with their beliefs? We investigate this question in a  model of opinion dynamics  and find surprising results.

There are several existing models of opinion dynamics on social networks. Many models are based on the classical voter model, in which agents have binary opinions (agreement or disagreement with a proposition) and then change their opinions randomly based on the opinion of a random neighbor or group of neighbors.\cite{holley1975ergodic,castellano2009statistical} Extensions of this model include making some of the voters ``stubborn,'' so that they do not change their opinions,\cite{yildiz2013binary,mobilia2003does} and assuming that   connections turn on  and off as agents change their opinions. \cite{del2017modeling,durrett2012graph,holme2006nonequilibrium,nickerson1998confirmation,zschaler2012early} One stubborn voter can sway the entire population\cite{mobilia2003does} and many stubborn voters can lead to opinion fluctuations.\cite{yildiz2013binary} Other models represent agents' beliefs with a number between $0$ and $1$, indicating the level of support for a proposition. These models can assume that agents are Bayesian,  meaning they combine data and prior beliefs correctly, but with the caveat that updates may be boundedly rational, i.e., Bayesian with constraints on resources used to compute. Therefore, many of these models institute approximations to Bayesian updates\cite{golub2010naive, acemoglu2011opinion}  to make the Bayesians boundedly rational. These approximations also reduce compute time. Generalizations to these models include adding cognitive biases,\cite{fernandes2023confirmation} fake news,\cite{azzimonti2018social} and stubborn voters.\cite{acemouglu2013opinion}

We use a Bayesian model that  represents an agent's belief as the likelihood that the agent assigns to a proposition being true given the evidence. The belief is updated according to a well-known nonlinear   relation,\cite{czegel2022bayes}  which includes an evidence term that the boundedly rational models lack. At each time step one agent updates their belief and then the agents connected to them update their beliefs. These updates are  interpreted as agents receiving some data such as an article that either supports a proposition to be true or disconfirms it. To implement confirmation  or anti-confirmation bias, the update relation  depends on whether the data agrees with the receiver's pre-existing belief. To implement in-group bias, the update  depends on whether or not the sender and receiver have similar beliefs.

If there is no cognitive bias, there is consensus. The introduction of cognitive biases leads to polarization, limit cycles, and even signatures of transient chaos. This  behavior  does not appear in  previous opinion dynamics models under similar conditions. In particular, we  find evidence for transient chaos\cite{lai2011transient} when agents attempt to counteract confirmation bias by employing anti-confirmation bias. Our simulations suggest that attempts to curtail the effects of cognitive biases on polarization and lack of use of ``wisdom of the crowd'' should be carefully monitored, because they may unexpectedly lead to the generation of chaotic behavior.

\section{The Model}\label{MandM}

In our model the agents are nodes in a network connected according to an adjacency matrix. Each node holds a belief $b_i$ that ranges from $0$ to $1$ and describes the probability that the agent assigns a proposition to be true. Beliefs are randomly initialized as independent uniform random numbers. The simulation then updates the agents' beliefs in the proposition by first randomly choosing an  agent in the network, giving the agent data, and having it update its belief.  Next, connected agents update their beliefs, which is interpreted  as these agents receiving data (a model of reposting on social media for example), and then neighbors of those agents update their beliefs and so forth.  When there are multiple agents to which one agent is connected, the beliefs of these  agents are updated in the order of their indices, though any order can be chosen. This process implies that the same agent updates their belief more than once. This updating process continues until the dynamical behavior of the network of agents is evident.

The adjacency matrix connects agents in a small-world network, which  reproduces certain features of real-world social networks, most notably  the existence of social hubs and the small number of ``hops'' required to go from one node to another.\cite{watts1998collective}  The parameters of the small-world network are chosen so that each node is connected to about a third of the nodes, although this choice is arbitrary. The number of connections per node is given by $K$. The adjacency matrix is an $N\times N$ matrix, where $N$ represents the number of agents in the   network. Each row of the matrix, $i$, represents an agent that can  influence  other nodes in the network, which means causing them to update their beliefs.  Each column of the matrix, $j$, represents the agent that is influenced by others. Nodes in the matrix $A$, denoted as $A_{ij}$, are filled with either zero or one, which represents whether or not agent $i$ can influence agent $j$. It does not need to be the case that $A_{ij} =  A_{ji}$, because it could be that one node influences another but that the influence is not returned, in the same way that someone might follow someone else on social media and not be followed back. The diagonal of the matrix, where $i=j$, is filled with zeros, because an agent cannot influence itself. The adjacency matrix is created before beliefs are updated. In the small world network, connections between nodes initially occur when they are neighbors, i.e. when their indices are within $1$ of each other. To create the small-world network, we then introduce some randomness into the network. We run through all nodes and break each connection with probability $\beta$, connecting the node to a new random node if that happens. Any value of $\beta$ can be chosen, and we used $\beta=0.5$.

The nodes have beliefs that are not binary, and  are represented by   $0 \leq b_i \leq 1$ that the proposition is true.\cite{golub2010naive,fernandes2023confirmation} In particular, we interpret $b_i$ as  $P(H=T|D)$, the probability that the proposition $H$  is true given all the data $D$ known so far. 
The Bayesian update of  beliefs is described in the following. 

We let $b_{i,t}$ be the belief of   agent $i$ after update $t$. According to Bayes' rule 
\begin{equation}
P_{i,t+1}(H=T|D_{t+1}) = \frac{P(D_{t+1}|H=T)P_{i,t}(H=T)}{P_i(D_{t+1})},\label{eq1}
\end{equation}
with
\begin{eqnarray}
P_i(D_{t+1}) &=& P(D_{t+1}|H=T)P_{i,t}(H=T)\nonumber \\
&&{} +P(D_{t+1}|H=F)P_{i,t}(H=F) \\
&=& P(D_{t+1}|H=T)b_{i,t} \nonumber \\
&&{} +P(D_{t+1}|H=F)(1-b_{i,t}). \label{eq:0}
\end{eqnarray}
After combining Eqs.~(\ref{eq1}) and (\ref{eq:0}) we see that the update equation depends only on the ratio $\gamma = P(D_{t+1}|H=F)/P(D_{t+1}|H=T)$, and not on the particular values of the likelihoods. These likelihoods are interpreted in a Bayesian framework rather than a frequentist one, meaning that the likelihood $P(D_{t+1}|H=F)$ is the agent's belief in how likely it is to have been shown the data $D_{t+1}$ given a model of the world that asserts that the hypothesis $H$ is false, $H=F$. There is no ground truth for these likelihoods. The likelihood $P(D_{t+1}|H=F)$ simply represents the agent's understanding of how likely it is that data $D_{t+1}$ would occur  given that the hypothesis is false. 
Note that $\gamma$  depends on both the receiver's belief $b_{i,t}$ and the sender's belief to implement various kinds of biases.
We substitute  the definition of $\gamma$ into Eq.~(\ref{eq:0}) and find
\begin{equation}
b_{i,t+1} = \frac{b_{i,t}}{b_{i,t}+\gamma (1-b_{i,t})}. \label{eq:1}
\end{equation}
The ratio  $\gamma$ used to update beliefs can be chosen somewhat arbitrarily, but  is interpreted as depending on what is actually true, the credibility that person $i$ assigns to the person sending data (in-group bias), and on the data itself (confirmation bias).  In our simulations, all $\gamma$ values are the same for all nodes. 

The key is that the probabilities $P(D_{t+1}|H=T)$ and $P(D_{t+1}|H=F)$  depend on the belief $b_{i,t}$. To include confirmation bias in the Bayesian updates, we allow the likelihood of the data to be higher when the article's proposition agrees with the person's original belief than when the data does not agree with the person's belief. This confirmation bias is implemented by having a different value of $\gamma$ if $b_{i,t}> 1/2$, than if  $b_{i,t}< 1/2$. Confirmation bias encompasses a range of other behaviors, but this  choice quantifies one aspect of confirmation bias. Anti-confirmation bias is just the opposite -- we interpret  that the likelihood to be lower than when the data does not agree with the agent's belief.

Bayesian agents can also have in-group bias. The key is that the probabilities $P(D_{t+1}|H=T)$ and $P(D_{t+1}|H=F)$ can now depend on the belief $b_{i,t}$ of the receiver and the belief of the sender $b_t'$. To add in-group bias into the updates, we have one value of $\gamma$ when $b_i >1/2$ and  $b_t'$ and $b_{i,t}$ are close to one another such as  $|b_t'-b_{i,t}|<0.1$, and a different value of $\gamma$ otherwise.

In our simulations, there is a ``ground truth'' $g$, i.e., whether or not the proposition is actually true. Complete agreement with ground truth is  given by either $b_i=1$ if the ground truth is that the proposition is true or $b_i=0$ if the proposition is false. There is an article (data $d$)  shown to every agent in turn that either supports or does not support the ground truth. Showing data to an agent is implemented in the simulation by updating the agent's belief $b_i$. The data can agree or  disagree with the  ground truth, in the same way that there is data that both supports and does not support the belief that  climate change exists. Every source is considered either credible ($c=1$) or not credible ($c=0$) by the receiver of the data. Even though there is ground truth, agents  do not have access to it. They merely have access to data that may or may not agree with the ground truth, as represented by the values of $\gamma$. The combination of all three -- the ground truth, the data, and the credibility of the source -- determines the likelihoods, and thus the value of $\gamma$. These parameters  are summarized in  Table~\ref{tab:my_label} and are  chosen   to illustrate a range of dynamical behavior. They have no grounding in reality, and it is  unclear   if they are accurate reflections of how people behave.

\begin{table}[h]
    \begin{tabular}{|p{1.2in}|p{1.2in}|p{1in}|p{3in}|} \hline 
         $P(D_{t+1}|H=F)$ & $P(D_{t+1}|H=T)$ & $\gamma$ & Description \\ \hline
         $0.85$ & $0.15$ & $0.176$ & The ground truth matches the data's proposition and the agent finds the source to be credible, $g=d$ and $c=1$. \\ \hline 
         $0.15$ & $0.85$ & $5.67$ & The ground truth  does not match the data's proposition and the  agent finds the source to be credible, $g\neq d$ and $c=1$.\\ \hline 
         $0.35$ & $0.65$ & $1.86$ & The ground truth  matches the data's proposition and the  agent finds the source to not be credible, $g=d$ and $c=0$.\\ \hline
         $0.65$ & $0.35$ &  $0.54$ & The ground truth does not match the data's proposition and the  agent finds the source to not be credible, $g\neq d$ and $c=0$.\\\hline
    \end{tabular}
    \caption{Description of simulation parameters with likelihoods, the ratio of the likelihoods $\gamma$, and a description of when they occur. Because the exact value of $\gamma$ matters, seemingly slight quantitative differences can  lead to qualitative differences in the simulation.}
    \label{tab:my_label}
\end{table}

We discuss three simulations   to study the effects of different combinations of  various biases on  belief behavior. Each simulation has different dynamical update rules which depend on the cognitive biases inherent in the model. In Sec.~\ref{sec:ModelA}, with no cognitive biases, the data is interpreted to be always credible. In Sec.~\ref{sec:ModelB}, with just confirmation bias, the data is interpreted to be credible when it agrees with the receiver's belief. In Sec.~\ref{sec:ModelC}, with confirmation  and in-group bias, the data is interpreted to be credible when it agrees with the receiver's belief and when the sender and receiver  belief scores are within $0.1$ of each other.  The algorithm for each simulation is listed in the respective section. 

\subsection{Characterizations of network behavior}

\subsubsection{Certainty and Polarization}

The certainty $C_t$  characterizes how ``stubborn'' agents are with their beliefs. Consider that after $t$ updates, the network is polled to gather information about the agents' updated beliefs. Consider the fraction of people who are $95\%$ or more certain of their belief. That is, they either believe that the proposition is less than $5\%$  ($b_{i,t} < 0.05$) likely to be true or believe that the proposition is more than $95\%$ likely to be true ($b_{i,t} > 0.95$). We call this fraction at  update $t$  the network's certainty $C_t$.

The  polarization $P_t$  characterizes the consensus within the network. Consider the fraction of agents in the network who believe that a proposition is true. If the fraction is $0$ or $1$, there is consensus and  all agents disagree or agree with the proposition. If  the fraction is not $0$ or $1$, there is polarization. We call this fraction at  update $t$  the network's polarization $P_t$. $b_i <1/2$ is considered to be equivalent to a belief of $0$, and  $b_i>1/2$ is considered to be equivalent to a belief of $1$.

\subsubsection{Lyapunov exponents and chaos}

Imagine monitoring the Euclidean distance between one trajectory and another trajectory that starts nearby. This  distance can grow or shrink exponentially. If it grows exponentially in the long time limit, there is  chaos; otherwise, there is  not. The rate of increase of the logarithm of the Euclidean distance is the first Lyapunov exponent. If there is an exponentially increasing distance, the first Lyapunov exponent is positive. There are $N$ Lyapunov exponents for a system with $N$ nodes. These exponents quantitatively characterize how an $N$-dimensional ellipsoid   grows or shrinks along a particular axis as it evolves.

Sometimes, there is  initial chaotic behavior and then the system falls into an attractor that is not chaotic and stays there. When this behavior occurs, the system is said to exhibit transient chaos. This behavior is  difficult to diagnose, but one diagnostic technique relies on finite-time Lyapunov exponents being positive and later turning negative.\cite{stefanski2010transient}

There are several methods for evaluating the Lyapunov exponents, some of which are more appropriate when we know the equations governing the dynamical system's evolution and some of which are more appropriate when  only  numerical data is available.\cite{sandri1996numerical} In our case, we know the governing equations, and  we can calculate successive Jacobians and use these to calculate the Lyapunov spectrum.

Suppose that node $i$ has its beliefs altered at update $t+1$. Without confirmation bias, the calculation of the Jacobian  is straightforward according to Eq.~\eqref{eq:1}:
\begin{equation}
\frac{\partial b_{j,t+1}}{\partial b_{k,t}} =
\begin{cases}
1 & (j=k,~j\neq i) \\
\dfrac{\partial b_{i,t+1}}{\partial b_{i,t}} = \dfrac{\gamma}{\left(b_{i,t}+\gamma(1-b_{i,t})\right)^2}& (j=k=i)   \\
0 & \text{otherwise}.\\
\end{cases} \label{eq:2}
\end{equation}
With confirmation bias, $\gamma$ depends discontinuously on $b_{i,t}$ at $b_{i,t}=1/2$, but there is zero probability of $b_{i,t}$ being exactly $1/2$. The Jacobian is  the matrix of these partial derivatives -- in other words, a diagonal matrix that is nearly the identity, but with one entry altered as in Eq.~\eqref{eq:2}. The Lyapunov spectrum is determined by multiplying successive Jacobians. We evolve the system and find the $QR$ decomposition of the resultant matrix, which is the decomposition of this matrix into an orthogonal and a triangular matrix. The logarithm of the diagonal entries of $R$, the upper triangular matrix,  grows linearly with the number of updates, and the growth rates are the Lyapunov exponents. Because the Jacobian here is a diagonal matrix, the diagonal entries are just the product of  $\partial b_{i,t+1}/\partial b_{i,t}$ over different times. We use $10^3$ updates total to compute the Lyapunov exponents, so they are finite-time Lyapunov exponents. We also evaluate them in Sec.~\ref{results} at $10^6$ updates to understand if we are finding transient chaos or chaos.

For the special case of a one-dimensional map $x_{t+1}=f(x_t)$, the Lyapunov exponent can be evaluated as
\begin{equation}
\lambda = \lim_{t\rightarrow\infty} \frac{1}{t}\sum_{t'=1}^t \log \left|\frac{\partial f}{\partial x}|_{x_{t'}}\right|.
\end{equation}
In our model $x = b_i$ and
\begin{equation}
\partial f/\partial x|_{x_{t'}} = \gamma/\left(b_{i,t'} + \gamma (1-b_{i,t'} )\right)^2.
\end{equation}

\section{\label{results}Results}

\subsection{Without any cognitive biases, all of society agrees}
\label{sec:ModelA}

\begin{figure}[!h]
    \begin{subfigure}[htbp]{0.45\textwidth}
        \begin{tikzpicture}
        \node[inner sep=0pt] (duck) at (0,0)
        {\includegraphics[width=\textwidth]{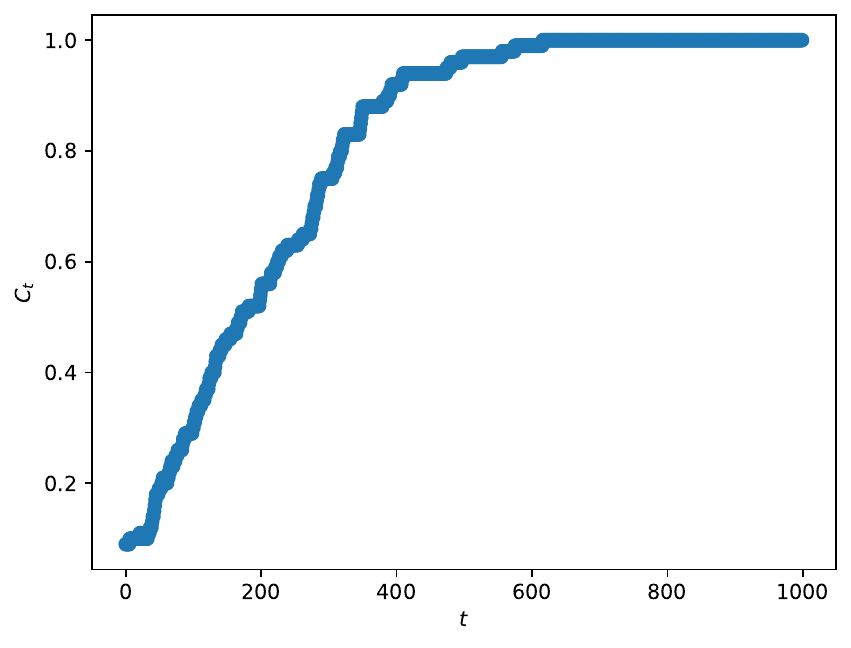}};
        \node[align=center,fill=white,draw] at (-2.4,2.1) {(a)};
        \end{tikzpicture}
    \end{subfigure}
    ~
    \begin{subfigure}[htbp]{0.45\textwidth}
        \begin{tikzpicture}
        \node[inner sep=0pt] (duck2) at (0,0)
        {\includegraphics[width=\textwidth]{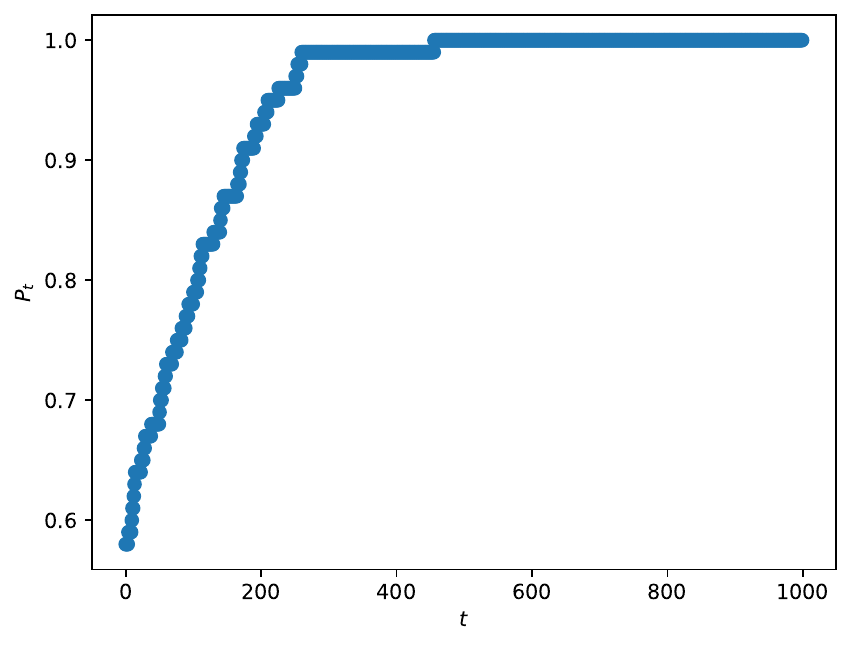}};
        \node[align=center,fill=white,draw] at (-2.4,2.1) {(b)};
        \end{tikzpicture}
    \end{subfigure}
    \caption{Confident consensus emerges without cognitive bias. 
(a) An example of the confidence $C_t$ versus $t$ for a system of 100 agents. We used $\gamma=0.176$, which could match a ground truth of the proposition being true, data that suggests that the proposition is true, and all credible agents. With more updates,  the agents become more confident in their beliefs. (b) An example from the same simulation of the polarization $P_t$ versus $t$. With more updates, the system reaches consensus.}
    \label{fig:confidence}
\end{figure}

To understand the behavior of this dynamical system, we first characterize the fixed points and their stability. Fixed points are points at which updates  do not change the state of the system. The stability of the fixed points is determined by changing the state of the system a little  from the fixed point and determining if  the system moves away or  toward the fixed point. 

This dynamical system has two fixed points  which we can identify by solving Eq.~\eqref{eq:1} for $b_{t+1}=b_t=b^*$:
\begin{equation}
b^{*} = \frac{b^*}{b^* + \gamma(1-b^*)} .
\end{equation}
The  resultant fixed points are
\begin{equation}
b^{*} = 0,1.
\end{equation}

We can categorize the stability of the fixed points by linearizing and analyzing the behavior of the nonlinear dynamical equations functions near a given fixed point by a Taylor expansion. We find that the stability of the fixed points  depends on $\gamma$.  To analyze the stability of the fixed points, we define $\delta b_t$ and $\delta b_{t+1}$ as
\begin{align}
\delta b_{t} &= b_t-b^{*}  \\
\delta b_{t+1} &= b_{t+1}-b^{*} .
\end{align}
We imagine that both $\delta b_t$ and $\delta b_{t+1}$ are small and the system is  near a fixed point. Because $b_{t+1} = f({b_t})$, we have
\begin{equation}
b^{*} + \delta b_{t+1} = f(b^{*}+\delta b_{t+1}) \approx f(b^{*}) + f'(b^{*}) \delta b_{t}
\end{equation}
At a fixed point, $b^{*} = f(b^{*})$, and hence
\begin{equation}
\delta b_{t+1}  \approx  f'(b^{*}) \delta b_{t}.
\end{equation}
For our system, we have
\begin{equation}
f'(b^{*}) = \frac{\gamma}{[(b^{*}-1) \gamma - b^{*}]^2} .
 \end{equation}
If $b^{*} =0$, $f'(b^*) = 1/\gamma$.
If $b^{*} =1$, $f'(b^*) = \gamma$.
To analyze the stability, note that if $|f'(b^*)| < 1$,
$|\delta b_{t}|$ decreases, and the fixed point is stable because further iterations push the system toward the fixed point.
However, if $|f'(b^*)| > 1$,
$|\delta b_{t}|$ grows and the fixed point is unstable and further iterations push the system away from the fixed point. 
Thus, we find that if $\gamma > 1$, $b^*=0$ is stable and $b^*=1$ is unstable. Conversely, we find that if $\gamma<1$, $b^*=0$ is unstable and $b^*=0$ is stable. If $\gamma=1$, $b_t$ does not change.

Our model has an important difference from previous belief networks for which beliefs converge to the eigenvector with eigenvalue $1$ of a transition matrix, and hence the convergent beliefs range between $0$ and $1$.\cite{golub2010naive,fernandes2023confirmation} Our belief network, with the inclusion of the evidence term, asserts that once an agent has made up its mind about something, it is difficult for it to change it. In particular, with no cognitive biases, every agent will  approach either $b_i=0$ or $b_i=1$ and will not oscillate from that position.

Our analytical analysis holds for one agent alone, and we may wonder if more complicated dynamics arise in the full network. In a higher dimensional space, we might see limit cycles or even chaos. But our simulations suggest that even in larger networks, e.g., $100$ agents, we typically see convergence to a fixed point in which all agents are   confident in their beliefs after $1000$ updates (see Fig.~\ref{fig:confidence}). This convergence occurs because in the opinion dynamics without any confirmation bias or in-group bias, the nodes act independently, and the network structure merely exists to order the nodes in terms of when they change their beliefs. Hence, with no cognitive biases, we will always see convergence to a consensus in which all agents are  confident of their beliefs.

Note from Table~\ref{tab:my_label} that there are two interpretations for  $\gamma < 1$  and two interpretations for  $\gamma > 1$. Thus, the consensus is not determined by the ground truth. To interpret it this way would require determining $\gamma$ from behavioral experiments.

\begin{algorithm}[H]
\caption{The simulation with no cognitive biases }\label{alg:cap1}
\begin{algorithmic}
\State Input $N$, $K$, $\beta$ and construct a small-world network, with $N=100$, $K=33$, and $\beta=0.5$ to reproduce the figures in Sec.~\ref{sec:ModelA}.
\State Randomly choose the ground truth $g$ and the data $d$ knowing that all sources are credible $c = 1$ because there are no biases.

\State Use the ground truth $g$, the data $d$, and the credibility of the source $c$ to find $\gamma$  in Table~\ref{tab:my_label}  \Comment{There is only one $\gamma$ value for all nodes.}
\State Randomly choose a starting node.
\While{There are nodes to update} 
\State Choose node to update if there is more than one node to update
\State Update node's belief using Eq.~\eqref{eq:1}. 
\State Update their Jacobian for calculation of Lyapunov exponents.
\EndWhile
\State Calculate Lyapunov exponents from the Jacobian.
\end{algorithmic}
\end{algorithm}

\subsection{With confirmation bias, polarization is possible}
\label{sec:ModelB}

As described in Sec.~\ref{MandM}, confirmation bias is implemented by having $\gamma$ depend on whether or not the node's belief agrees with the data. As before, nodes act independently, linked only by when they update. As such, we can analyze the network by analyzing the one-dimensional map:
\begin{equation}
b_{t+1} =
\begin{cases} \dfrac{b_t}{b_t+\gamma_\ell (1-b_t)} & b_t<\frac{1}{2} \\ \dfrac{b_t}{b_t+\gamma_h (1-b_t)} & b_t \geq \frac{1}{2} \end{cases} \label{eq:3}
\end{equation}
for some $\gamma_\ell$ ($b_t \leq \frac{1}{2}$) and $\gamma_h$ ($b_t \geq \frac{1}{2}$). We interpret our simulations as if there is  a ground truth and a data type which agrees with the ground truth or does not agree with ground truth, and the credibility of the source  depends entirely on whether or not $b_t$ agrees with the data (agrees if $b_t > \frac{1}{2}$ and the data supports the proposition being true or if $b_t<\frac{1}{2}$ and the data supports the proposition being false). It is this credibility that influences $\gamma$ to switch values depending on $b_t$. 
Dynamically, $\gamma_\ell$ and $\gamma_h$ determine if $0$ and $1$ are stable or unstable fixed points. Three cases arise:
\begin{itemize}
\item $b^*=0$ or $b^*=1$ is a stable fixed point, and the other fixed point is unstable, under both possible updates. These  outcomes  occur if either $\gamma_\ell$ and $\gamma_h>1$ or if $\gamma_\ell$ and $\gamma_h<1$ respectively.
\item For $\gamma_\ell>1$ and $\gamma_h<1$ both fixed points are  stable, and we have bistability. In this case $b_t$  approaches  either $0$ or $1$ depending on if $b_t$ is initially   less than $1/2$ or greater than $1/2$.
\item For $\gamma_\ell<1$ and $\gamma_h>1$ both fixed points are  unstable, and $b_t$  oscillates around $1/2$. The exact attractor depends on the initial belief $b_0$, i.e., a slightly different value of $b_0$ leads to a slightly shifted attractor.
\end{itemize}

We ignore the case for which $\gamma_\ell$ or $\gamma_h$ is exactly equal to $1$.
These three cases correspond to wildly different scenarios. The first case corresponds to the situation in which the data speaks for itself, even with confirmation bias, and agents converge on believing that the data represents ground truth. The second case corresponds to the situation in which confirmation bias  holds completely, and the article's proposition is largely irrelevant. Final beliefs are based on initial beliefs entirely, and so the system can become polarized. (Everyone will be very certain and fixed in their beliefs, which will just mirror their initial beliefs.) The third   case, which is the most dynamically interesting, corresponds to a weird phenomenon in which the opposite of confirmation bias happens -- when one's belief is that the proposition is false, we evaluate data to say that the proposition is true, and vice versa.

\begin{figure}[!h]
    \begin{subfigure}[htbp]{0.45\textwidth}
        \begin{tikzpicture}
        \node[inner sep=0pt] (duck) at (0,0)
        {\includegraphics[width=\textwidth]{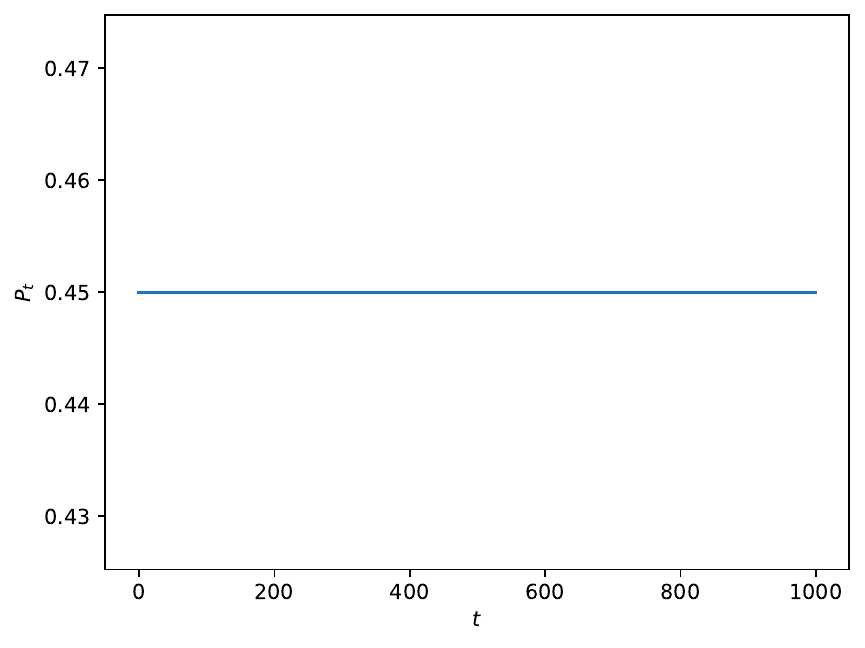}};
        \node[align=center,fill=white,draw] at (-2.4,2.1) {(a)};
        \end{tikzpicture}
    \end{subfigure}
    ~
    \begin{subfigure}[htbp]{0.45\textwidth}
        \begin{tikzpicture}
        \node[inner sep=0pt] (duck2) at (0,0)
        {\includegraphics[width=\textwidth]{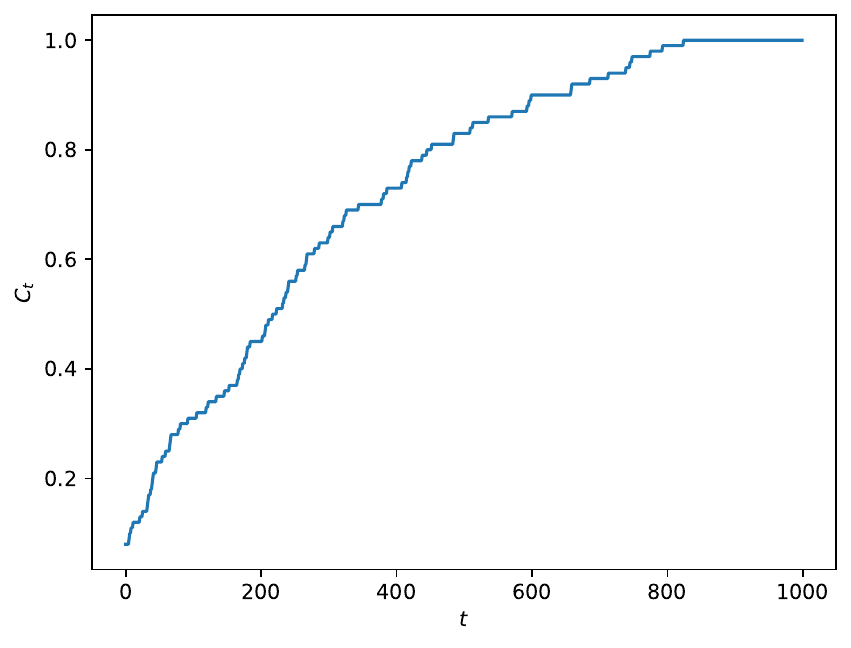}};
        \node[align=center,fill=white,draw] at (-2.4,2.1) {(b)};
        \end{tikzpicture}
    \end{subfigure}
    ~
    \begin{subfigure}[htbp]{0.45\textwidth}
        \begin{tikzpicture}
        \node[inner sep=0pt] (duck3) at (0,0)
        {\includegraphics[width=\textwidth]{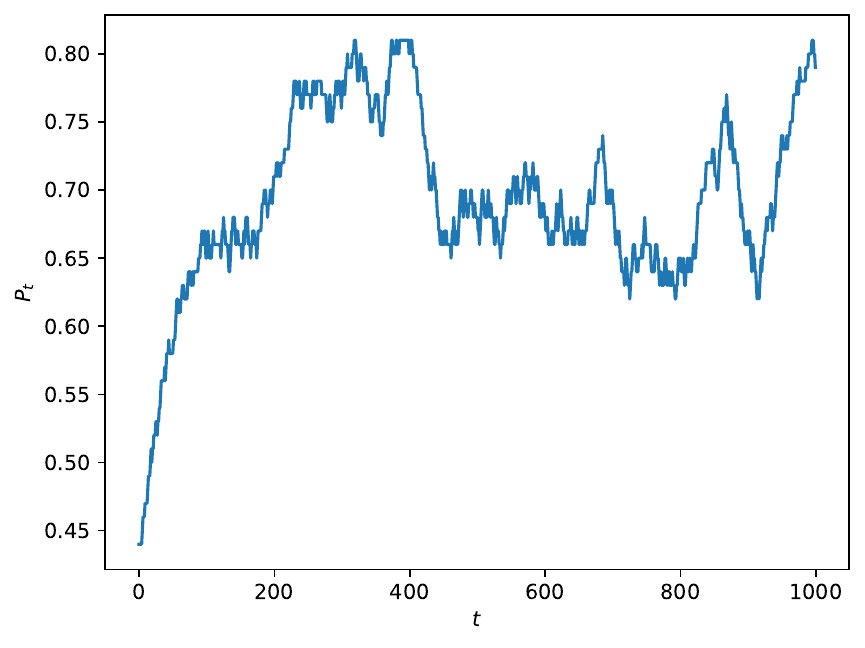}};
        \node[align=center,fill=white,draw] at (-2.4,2.1) {(c)};
        \end{tikzpicture}
    \end{subfigure}
    ~
    \begin{subfigure}[htbp]{0.45\textwidth}
        \begin{tikzpicture}
        \node[inner sep=0pt] (duck4) at (0,0)
        {\includegraphics[width=\textwidth]{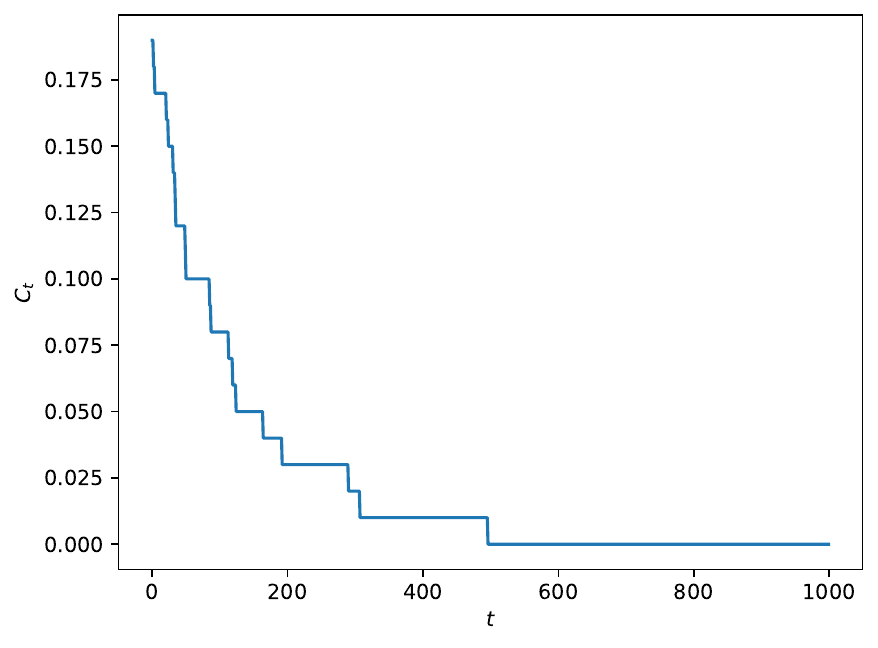}};
        \node[align=center,fill=white,draw] at (-2.4,2.1) {(d)};
        \end{tikzpicture}
    \end{subfigure}
    \caption{Polarization and belief cycling can both occur, with different $\gamma_\ell$ and $\gamma_h$ values for the two behaviors as listed below with $10$ agents.  
(a) and (b) We set $\gamma_\ell=5.67$ and $\gamma_h=0.54$ which could result from the ground truth  being that the proposition is true and the data disagreeing with this ground truth. The certainty $C_t$  and polarization $P_t$  versus the number of updates $t$ shows that  there is convergence to a fixed point with bistability.  (c) and (d) We set  $\gamma_\ell=0.176$ and $\gamma_h=1.86$ which could result from the ground truth  being that the proposition is false and the data agreeing with this ground truth.  The certainty $C_t$  and polarization $P_t$ versus  $t$ shows that there is belief cycling.}
    \label{fig:belief_cycling}
\end{figure}
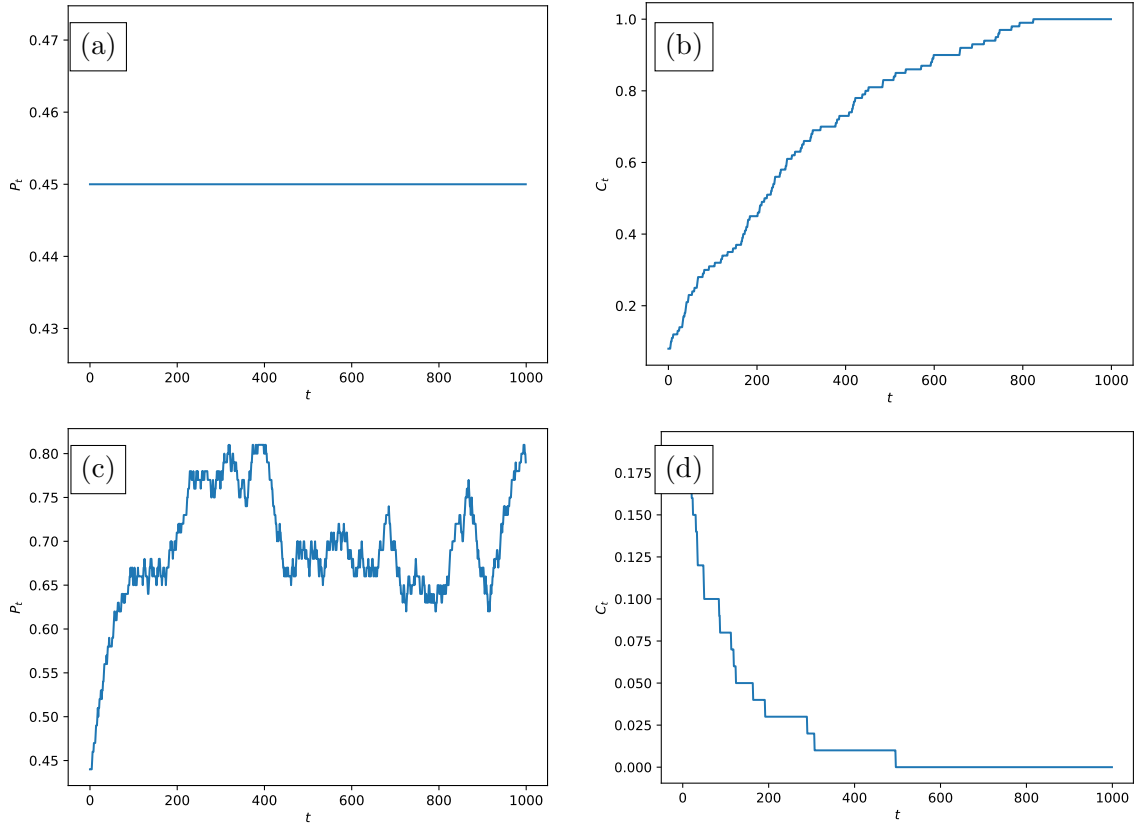

With the parameter settings listed in Table~\ref{tab:my_label}, polarization and belief cycling are both possible. In these simulations, all network nodes are identical in how they evolve their beliefs and so have the same value of $\gamma_\ell$ and $\gamma_h$ (see Fig.~\ref{fig:belief_cycling}).  Polarization and belief cycling can both occur. 
When the ground truth  is that the proposition is true and the data disagrees with this ground truth, meaning that  $\gamma_\ell=5.67$ and $\gamma_h=0.54$, the certainty $C_t$ grows  and polarization $P_t$ levels out as the number of updates $t$ increases, implying convergence to a fixed point with bistability.  When the ground truth  is that the proposition is false and the data agrees with this ground truth, which means that the parameters here are $\gamma_\ell=0.176$ and $\gamma_h=1.86$, the certainty $C_t$ decreases and the polarization $P_t$ fluctuates, implying belief cycling.

\begin{algorithm}[H]
\caption{The simulation with confirmation bias only}\label{alg:cap2}
\begin{algorithmic}
\State Choose $N$, $K$, $\beta$ and construct a small-world network, with $N=10$, $K=3$, and $\beta=0.5$ to reproduce the figures in Sec.~\ref{sec:ModelB}.
\State Randomly choose the ground truth $g$ and the data $d$, with source credibility $c$ determined later, to determine $\gamma_{\ell}$ and $\gamma_h$.
\State Choose a starting node.
\While{There are nodes to update}
\State Choose node to update if there is more than one node to update
\If {Data agrees with the belief of the receiver}
\State Credibility is high, $c=1$.
\State Find $\gamma$ value from Table~\ref{tab:my_label} using $g,~d,~c$.
\Else
\State Credibility is low, $c=0$.
\State Find $\gamma$ value from Table~\ref{tab:my_label} using $g,~d,~c$.
\EndIf
\State Update each receiver node's belief using Eq.~\eqref{eq:1}.
\State Update their Jacobian for calculation of Lyapunov exponents.
\EndWhile
\State Calculate Lyapunov exponents from the Jacobian.
\end{algorithmic}
\end{algorithm}

\subsection{With anti-confirmation bias and in-group bias, transient chaos can occur}
\label{sec:ModelC}
When we allow for both anti-confirmation bias and in-group bias so that the nodes are coupled and their evolution is nonlinear, we find that signatures of transient chaos can emerge. We can use the previous analysis as a guide. For instance, if the ground truth is that the proposition is true, then the $\gamma$ values implied (see Sec.~\ref{MandM}) correspond to polarization based on the analysis in Sec.~\ref{sec:ModelB}, but consensus may occur depending on initial conditions. If the ground truth is that the proposition is false, the $\gamma$ values according to Table~\ref{tab:my_label} correspond to belief cycling. We stress that this analysis is only a guide because of the effect of in-group bias. 
Aspects of the prior analyses still hold. With anti-confirmation bias, there are a wealth of unstable fixed points given by $b_i=0$ or $b_i=1$ for all $i$ and a wealth of stable limit cycles which can be seen more clearly by simulating for  $10^6$ iterations. After what looks like it could be transient chaos, the exact stable limit cycle  depends sensitively on the initial conditions. The dissipation rate (the sum of the Lyapunov exponents) in the  long time limit  seems to be given by at most $-7\times 10^{-4}$ per active node and is less negative if some of the nodes are disconnected because the small-world network fails to produce a fully-connected network. When the simulation is run for a small number of updates ($10^2$), the positive finite-time Lyapunov exponents can reach $\sim 8\times 10^{-3}$. This finite-time Lyapunov exponent is consistent  with the doubling time seen on plots of the Euclidean distance as a function of time (not shown here). As such, we take these small but positive finite-time Lyapunov exponents to be meaningful. Note that the positive finite-time Lyapunov exponents at $10^4$ iterations (which are maximally around $10^{-4}$) are likely to be smaller than the Lyapunov exponents at $10^2$ iterations if there is apparent transient chaos which crosses over to a stable attractor, because the apparent transient chaos is perhaps being averaged in with the negative contributions to the finite-time Lyapunov exponent of the stable limit cycle.

\begin{figure}[!h]
    \begin{subfigure}[htbp]{0.45\textwidth}
        \begin{tikzpicture}
        \node[inner sep=0pt] (duck) at (0,0)
        {\includegraphics[width=\textwidth]{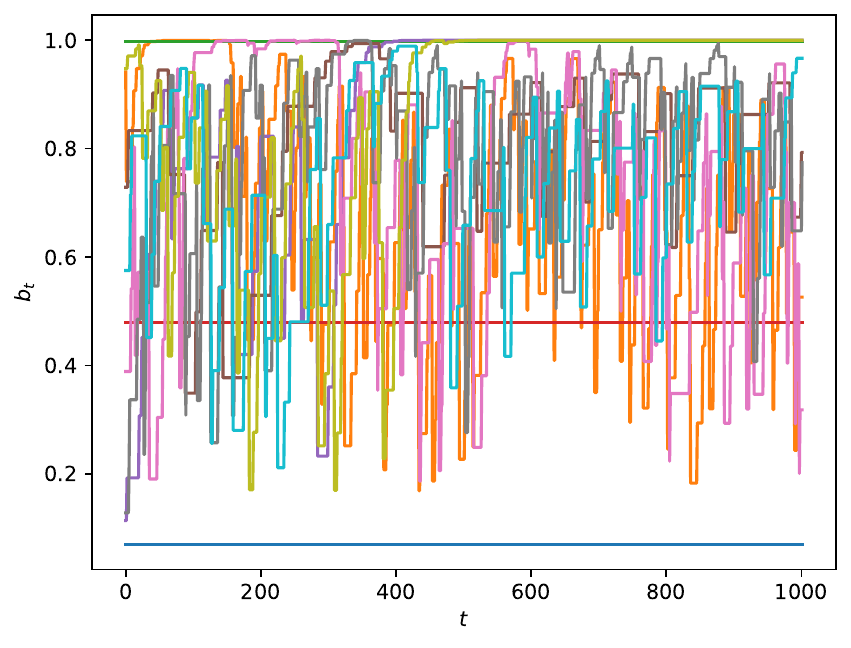}};
        \node[align=center,fill=white,draw] at (-2.4,2.1) {(a)};
        \end{tikzpicture}
    \end{subfigure}
    ~
    \begin{subfigure}[htbp]{0.45\textwidth}
        \begin{tikzpicture}
        \node[inner sep=0pt] (duck2) at (0,0)
        {\includegraphics[width=\textwidth]{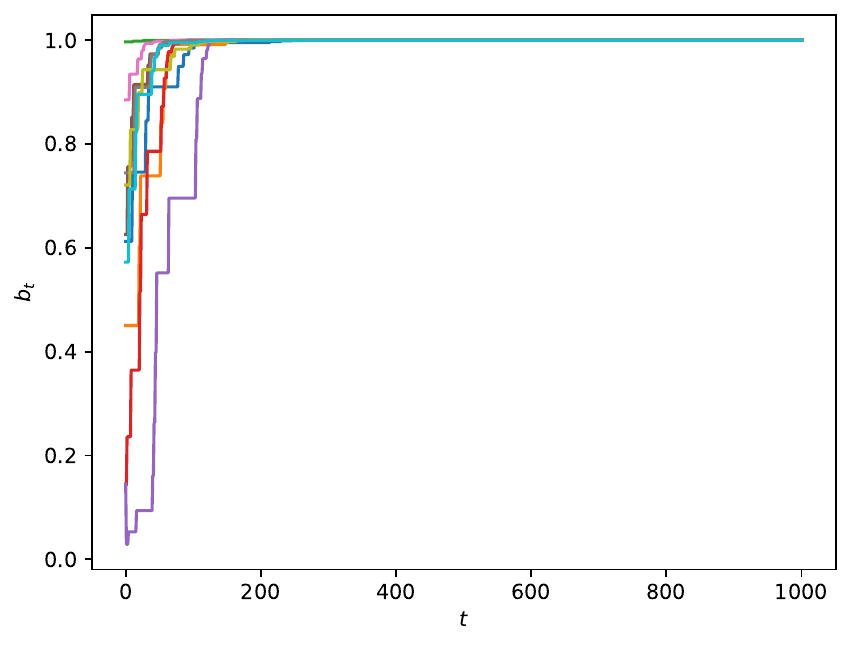}};
        \node[align=center,fill=white,draw] at (-2.4,2.1) {(b)};
        \end{tikzpicture}
    \end{subfigure}
    \caption{(Color online) Transient chaos or consensus for $10$ agents.  (a) Transient chaos can emerge when there is both anti-confirmation  and in-group bias. The ground truth in this simulation is that the proposition is false, but the data being shown to agents disagrees, which leads to $\gamma=5.67$ when the data is credible and $\gamma=0.54$ when the data is not  credible. Without in-group bias, these parameter values would lead to a limit cycle. The beliefs as a function of time suggest transient chaos, with positive finite-time Lyapunov exponents of $2.6\times 10^{-4}$ and $4.5\times 10^{-4}$. The rest of the finite-time Lyapunov exponents are negative.
Each node is represented by a different color.  (b)  The beliefs as a function of time show consensus when the ground truth is that the proposition is true and the data disagrees, implying that $\gamma=5.67$ when the data is considered credible and $\gamma=0.54$ when the data is not  credible. These parameter values would lead to bistability and polarization without in-group bias. There are no positive finite-time Lyapunov exponents.}
    \label{fig:chaos}
\end{figure}

Transient chaos is suggested in  Fig.~\ref{fig:chaos}  by the complicated aperiodic evolution of some of the beliefs and by some small but positive finite-time Lyapunov exponents. The Lyapunov spectrum for this system and elements of the Jacobian are surprisingly small in magnitude, so that the small but positive Lyapunov exponents may actually be meaningful. We also can see consensus, polarization, and belief cycling depending on the initial conditions and parameters, as we did  with models that do not include in-group bias, with accompanying negative Lyapunov exponents  (see Fig.~\ref{fig:chaos}).  

For $10^6$  updates, the finite-time Lyapunov exponents are negative and for a $10$-node network  cluster around $-7\times 10^{-4}$. The system remains at a stable fixed point or a stable limit cycle, and so we cannot report finding chaos (in contrast to transient chaos) for these network dynamics. Because we did not test all initial conditions and because there is stochasticity in the governing equations, we cannot definitively rule out chaos for some set of initial conditions and for some values of $\gamma$ (see Sec.~\ref{MandM}).

In short, a combination of anti-confirmation bias, which leads to unstable fixed points at beliefs that indicate certainty, and in-group bias, which leads to the evolution of one node's beliefs depending on a non-differentiable and nonlinear way on other nodes' beliefs, yield complex opinion dynamics with sometimes positive finite-time Lyapunov exponents. The attractors to which the system converges depend  strongly on the initial conditions and the values of $\gamma$ which depend on the ground truth, if the data supports the ground truth, and the closeness of the receiver's and sender's beliefs. 

\begin{algorithm}[H]
\caption{The simulation with confirmation bias and in-group bias}\label{alg:cap3}
\begin{algorithmic}
\State Choose $N$, $K$, $\beta$ and construct a small-world network, with $N=10$, $K=3$, and $\beta=0.5$ to reproduce the figures in Sec.~\ref{sec:ModelC}.
\State Randomly choose the ground truth $g$ and the data $d$, with source credibility $c$ determined later, to determine $\gamma$.
\State Choose a starting node.
\While{There are nodes to update}
\State Choose node to update if there is more than one node to update
\If {Data agrees with the belief of the receiver and the source node's belief is within $0.1$ of the receiver node's belief}
\State Credibility is high, $c=1$.
\State Find $\gamma$ value from Table~\ref{tab:my_label} using $g,~d,~c$.
\Else
\State Credibility is low, $c=0$.
\State Find $\gamma$ value from Table~\ref{tab:my_label}  using $g,~d,~c$.
\EndIf
\State Update each receiver node's belief using Eq.~\eqref{eq:1}. 
\State Update their Jacobian for calculation of the Lyapunov exponents.
\EndWhile
\State Calculate Lyapunov exponents from the Jacobian.
\end{algorithmic}
\end{algorithm}

\section{Conclusion}

We have found that opinion dynamics can show richer dynamical behavior than has been  found previously.\cite{strogatz2018nonlinear,lai2011transient} In addition to polarization, belief networks with individuals that  attempt to overcome their confirmation bias by exhibiting anti-confirmation bias, but who show in-group bias, can show belief cycling (limit cycles) and apparent transient chaos. This behavior suggests that opinion dynamics are likely to be chaotic when all cognitive biases and attempts to counteract cognitive biases are taken into account, and when there are enough nodes for the cognitive biases to take hold.

Surprisingly, we  also found  qualitative results that are opposite to that of Ref.~\onlinecite{ngampruetikorn2016bias} in that confirmation bias can lead to polarization but lack of confirmation bias cannot. The dynamics used here    assumes that the agents are Bayesian, while the opinion dynamics of Ref.~\onlinecite{ngampruetikorn2016bias} and related references are somewhat ad hoc. Behavioral experiments will be needed to reveal whether or not confirmation bias leads to polarization or stabilizes consensus, with the following caveat.

The reason for  opinion fluctuations is likely related to the presence of stubborn agents who stick to their opinion once formed no matter what. In our simulations, stubborn agents are formed naturally; in prior work, they were hard-coded\cite{acemoglu2011opinion,acemouglu2013opinion,mobilia2003does} and led to fluctuating opinions. Interestingly, the mechanism of the opinion fluctuations differs. In prior work, the stubborn agents alone were sufficient to induce fluctuations. In our simulations, cognitive biases are required. The natural Bayesian updates of priors to posteriors are not enough. Hence, the presence of stubborn voters is not enough to lead to fluctuating opinions when fully Bayesian networks are simulated. Boundedly rational update rules are required.

We found that   reducing confirmation biases which produce polarization can lead to chaotic behavior in  simple belief networks. Attempts to control chaos with timed injection of new data by changing $\gamma$ during the simulation might become important for controlling opinion dynamics on the macroscopic level even after confirmation bias is reduced or actively counteracted. Simulations  may be useful  for understanding the surprising effects of attempts to counteract cognitive biases, and more unexpected effects may result from more complex simulations that include a rewiring of the social network, as do some statistical physics models of opinion dynamics. \cite{del2017modeling,durrett2012graph,holme2006nonequilibrium,nickerson1998confirmation,zschaler2012early}

Finally, we discuss the occurrence of aapparent transient chaos in contrast   to chaos. Assuming the signatures of transient chaos are indeed indicative of transient chaos, does  it matter if there is transient chaos? Do  we  care only about what happens in the asymptotic limit? If transient chaos governs the initial opinion dynamics once a question is posed in a democracy, its presence is just as important for policy makers who hope to manipulate public opinion as the asymptotic properties of the system. Also, there are other ways to characterize transient chaos, but in the high-dimensional systems considered, they seemed impractical. For instance, it is impossible to measure escape rates\cite{lai2011transient} because we do not have a good sense of the region in which the attractor existed. Future efforts could concentrate on assuming a changing social network, allowing for many kinds of data to be shown to agents in the network, and also for a more complete characterization of the transient chaos to obtain a better understanding of how to manipulate it so that consensus is achieved and information is integrated by the social network. In addition, admitting more than a binary true/false might lead to chaos without even having confirmation biases or any sense of bounded rationality.\cite{sato2003coupled}

In summary, the novel ingredients to our analysis are that we have developed a cognitive science model based on a Bayesian framework and analyzed it in a dynamical systems framework. In the process, we calculated finite-time Lyapunov exponents in an opinion dynamics model for the first time. We hope that some of the other models in the literature will be analyzed similarly for fixed points, limit cycles, and transient chaos or chaos.

\section{Suggested Problems}
\begin{itemize}
\item Confirm the analysis of Sec.~\ref{sec:ModelB} with fixed point arguments. How can we obtain polarization when the results of Sec.~\ref{sec:ModelA} indicated consensus?
\item  Program the three simulations and find evidence for limit cycles as in Sec.~\ref{sec:ModelB} and signatures of transient chaos as in Sec.~\ref{sec:ModelC}.
\item Choose your own parameters for $N$, $K$, $\beta$, and the likelihoods and see what changes. Do the fixed point arguments of Sec.~\ref{sec:ModelA} and Sec.~\ref{sec:ModelB} remain valid?
\end{itemize}

\begin{acknowledgments}
This study was supported by the U.S.\ Air Force Office for Scientific Research, Grant Number FA9550-19-1-0411. Thanks go to Jim Crutchfield, Joshua Garland, and Mirta Galesic for edifying and illuminating conversations.
\end{acknowledgments}

\bibliography{chaos}

\end{document}